\documentclass[12pt]{article}

\usepackage[russian]{babel}% Русификация
\usepackage{graphicx,graphics}
\begin{document}
\begin{center}
{\large\bf Explanation of velocities distribution in the galaxies without the dark matter} \vskip 0.3 true in
{\large V. N. Borodikhin} \vskip 0.3 true in {\it
Omsk State University, pr. Mira 55a, Omsk, Russia.}
\end{center}
%\date{\today}
\begin{abstract}%
In this work, we submit the explanation of spectra of the rotary curves of galaxies on the basis of a vector
 theory of gravitation without regard to the dark matter hypothesis. We study the approximation of rotary
  curves in case of existing of the medium constant cyclic field by the example of a number of galaxies.
   Besides, we obtain the curves of distribution of cyclic fields in the galaxies, wherein the theoretical
    and experimental data is equal.
\end{abstract}

\vskip 0.2 true in e-mail: borodikhin@inbox.ru

\section{Introduction}

As it is well known, there is a number of observed effects, particularly the rotary curves of
 the galaxies, for which explanation the concept of the hidden mass, so called dark matter, is introduced  \cite{1,2}.
But there is no any direct experimental evidence of existing of such an exotic matter. In particular,
 last observations of the heavy WIMP-particles in the framework of the project XENON100 was
  negative \cite{3}. In addition to the dark matter, there are alternative explanations of these
   effects connected with the modification of the Newton's law
\cite{4,5}.

In this work, we submit the explanation of spectra of the rotary curves of galaxies on the basis of a
 vector theory of gravitation
\cite{6}. In the framework of this theory, we assume the existence of a vector cyclic field connected
 with a gravitation field in much the same manner as the magnetic field is collated with the electrical one.
  This cyclic field supposedly determines the anomalous behavior of rotary curves.

In this article, we study the approximation of rotary curves in case of existing of the medium
 constant cyclic field by the example of a number of different types of galaxies. Besides, we obtain
  the exact curves of distribution of cyclic fields in relation to the distance to the galactic centers,
   wherein the theoretical and test information is practically equal.

\section{The general model}

We will connect the gravitational field with the 4-potential $A^i=(\varphi, c\vec A)$, where
$\varphi$- is the usual scalar potential and $\vec A$ is a vector potential, and $c$ is the speed of light.
The Lagrangian of the gravitational field with
account for matter has the form

\begin{equation}\label{4}
\L=-A_ij^i+\frac{1}{16\pi\gamma}G_{ik}G^{ik},
\end{equation}

where $\gamma$ is the gravitation constant,
$j^i=\mu\frac{1}{c}\frac{dx^i}{dt}$ is
the mass current density vector
, $\mu$- is the mass density
of bodies, and
$G_{ik}=\frac{\partial A_k}{\partial x^i}-
\frac{\partial A_i}{\partial x^k} $ is the antisymmetric tensor of the gravitation field.

The first term describes interaction of the field and
matter, the second one characterizes the field without
particles.
As a result, we get the gravitational field equations

\begin{equation}\label{5}
\frac{\partial G^{ik}}{\partial x^k}=4\pi\gamma j^i.
\end{equation}

In the stationary case, from (\ref{5}) we obtain an equation
for the scalar potential:

\begin{equation}\label{7}
\triangle\varphi=4\pi\gamma\mu.
\end{equation}

The solution of (\ref{7}) has the form

\begin{equation}\label{8}
\varphi=-\gamma\int\frac{\mu}{r}dV.
\end{equation}

The potential of a single particle of mass m $\varphi=-\frac{\gamma m}{r}$.
Consequently the force acting in this field on
another particle of mass $m^{\prime}$ is

\begin{equation}\label{9}
F=-\frac{\gamma m m^{\prime}}{r^2},
\end{equation}

(\ref{9}) which is the Newton law of gravity. The negative sign
in this expression is caused by the positive sign of the
second term in the Lagrangian (\ref{5}), contrary to the
electromagnetic field Lagrangian.

Let us consider the field of the vector potential created
by matter particles performing motion in a finite
region od space with finite momenta. The motion of
this kind can be considered to be stationary. Let us
write down an equation for the time-averaged vector
field, depending only on spatial variables.

From (\ref{5}) we obtain:

\begin{equation}\label{10}
\triangle\overline{\vec A}=4\pi\gamma\overline{\vec j} ,
\end{equation}

whence it follows

\begin{equation}\label{11}
\overline{\vec A}= -\frac{\gamma}{c^2}\int\frac{\overline{\vec j}}{r}dV .
\end{equation}

The overline denotes a time average. This field can be
called cyclic. The field induction is

\begin{equation}\label{12}
\overline{\vec C}=rot\overline{\vec A}= -\gamma\int\frac{[\overline{\vec j}\vec r]}{r^3}dV=-\gamma,
\frac{[\overline{\vec p}\vec r]}{r^3},
\end{equation}

where $\vec p$ is the particle momentum and the square
brackets denote a vector product.

Thus two moving particles experience (in addition
to the mutual gravitational attraction) a cyclic force.
The latter can be attractive or repulsive, depending on
the relative direction of the particle velocities.

\section{Galaxy Rotation Curves. Approximation.}

Let us assume that apart from the ordinary gravity there is the said cyclic field all over
 the galaxy. We can make a simplifying assumption that this field has medium constant induction
  across the galaxy. There are probably sure changes of value of induction of a cyclic field
   describing by some function of distribution in each specific galaxy, but for generality of the
    arguments we can neglect this. Thus, let us write the condition of equality of centrifugal
     and centripetal forces:

\begin{equation}\label{30}
\frac{v^2}{r}=\frac{\gamma M(r)}{r^2}+vC
\end{equation}

where $M(r)$ - is the matter mass inside the orbit of the radius $r$, $C$ - is the constant induction of a cyclic field,
$\gamma$ - is the gravitational constant.
This is a quadratic equitation relative to velocity $v$. Solving it we obtain:

\begin{equation}\label{31}
v=\sqrt{\frac{\gamma M(r)}{r}}\sqrt{1+\frac{C^2r^3}{4\gamma M(r)}}+\frac{Cr}{2}
\end{equation}

Distribution of the ordinary matter in the sphere of the radius  $r$ takes the form:

\begin{equation}\label{32}
M(r)=4\pi\int\limits_{0}^{r}\rho(r^{'})r^{'2}dr^{'}
\end{equation}

where $\rho(r)$ is the matter density.

A simple model for $M(r)$ \cite{5}

\begin{equation}\label{33}
M(r)=M\Bigl(\frac{r}{r+r_c}\Bigr)^{3\beta}
\end{equation}

where

\begin{eqnarray}\label{34}
\begin{array}{lcl}
    \beta & =& \left\{ \begin{array}{ll}
                1 \  \ for  \  \ HSB \  \ galaxies, \\
                2 \  \ for \  \  LSB / Dwarf \  \ galaxies.  \\
             \end{array}
          \right.
\end{array}
\end{eqnarray}

If $r\ll r_c$ the influence of a gravitation field is considerably superior to the
 influence of a cyclic field, and we can neglect it in the 0 - order using just the Newton's theory.

The total mass of the galaxy  $M$ we determine as

\begin{equation}\label{35}
M = M_{stars} +M_{HI} +M_{f} ,
\end{equation}

Here, $M_{stars}$,$M_{HI}$ , and $M_{f}$ denote the visible mass,
the mass of neutral hydrogen,  and the mass from the
skew field energy density, respectively.

In fig.1. - fig.12. there are graphs of velocity distribution in the galaxies of different
 types. White circles are velocity values under study, black triangles are Newtonian
  approximations with regard to  (\ref{33}), black circles are the approximations
   with regard to the medium constant cyclic field $C$ obtained by the least square
    method. The data of velocity distribution is taken from  \cite{7,8}.

\begin{figure}
 \hspace{-1.5cm}%   \centering
 \includegraphics[width=9cm,height=6.5cm]{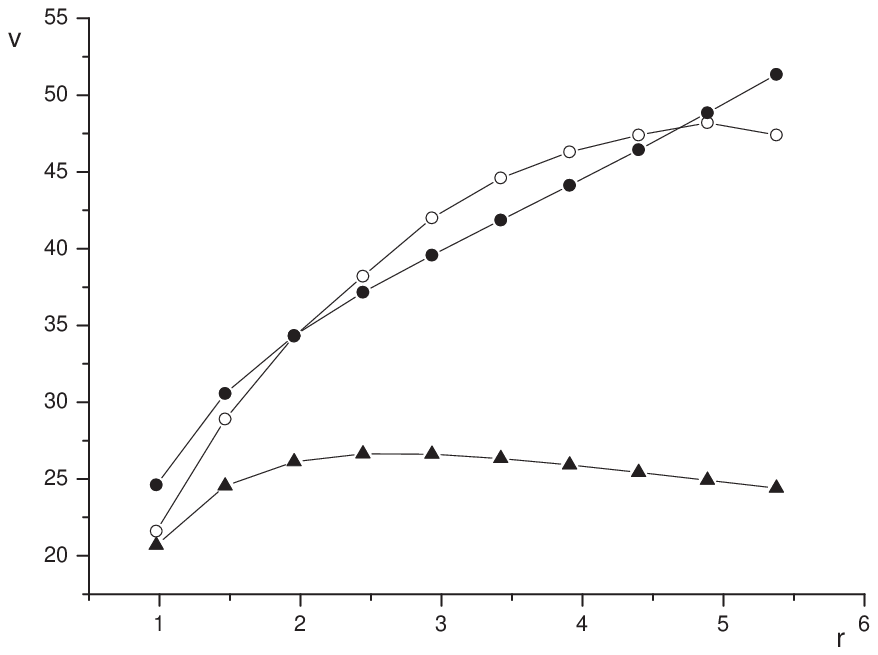}
 %\hspace{-2cm}
%\caption{Ddo154; C=1.6} % Подпись рисунка
\includegraphics[width=9cm,height=6.5cm]{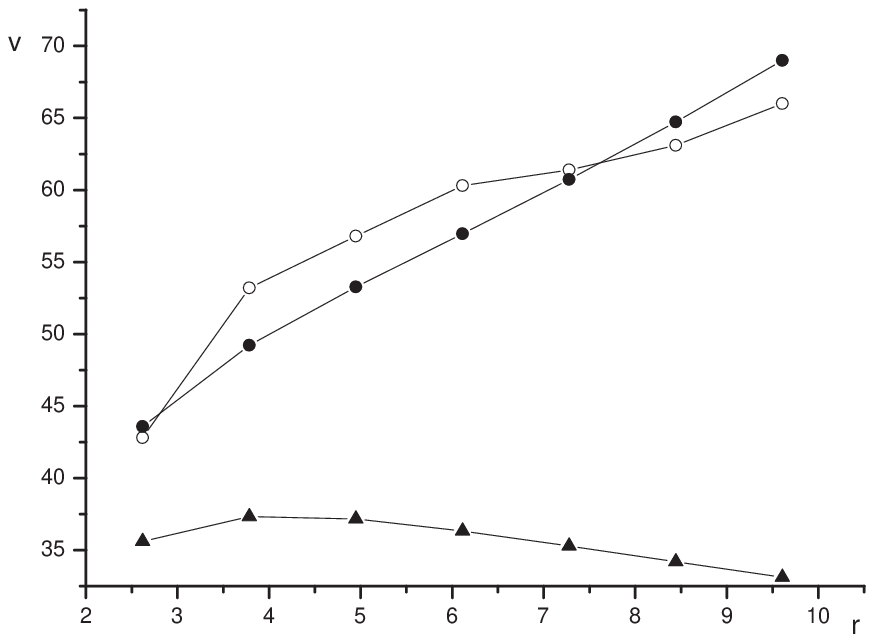}
%\caption{Ddo154; C=1.6} % Подпись рисунка
%\label{101} % Метка для ссылки на рисунок.
Fig.1. DDO 154; $\langle C\rangle=2.4*10^{-16}$.  \  \  \  \  \  \  \  \  \  \  \  \  \  \
Fig.2. DDO 170; $\langle C\rangle=1.7945*10^{-16}$.
\end{figure}

\begin{figure}
 \hspace{-1.5cm}%   \centering
 \includegraphics[width=9cm,height=6.5cm]{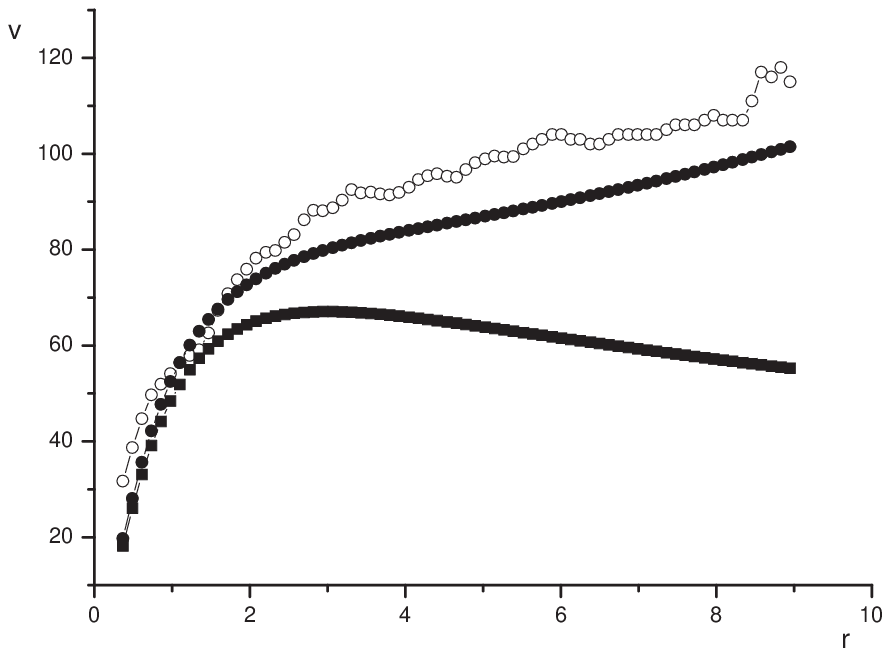}
 %\hspace{-2cm}
%\caption{Ddo154; C=1.6} % Подпись рисунка
\includegraphics[width=9cm,height=6.5cm]{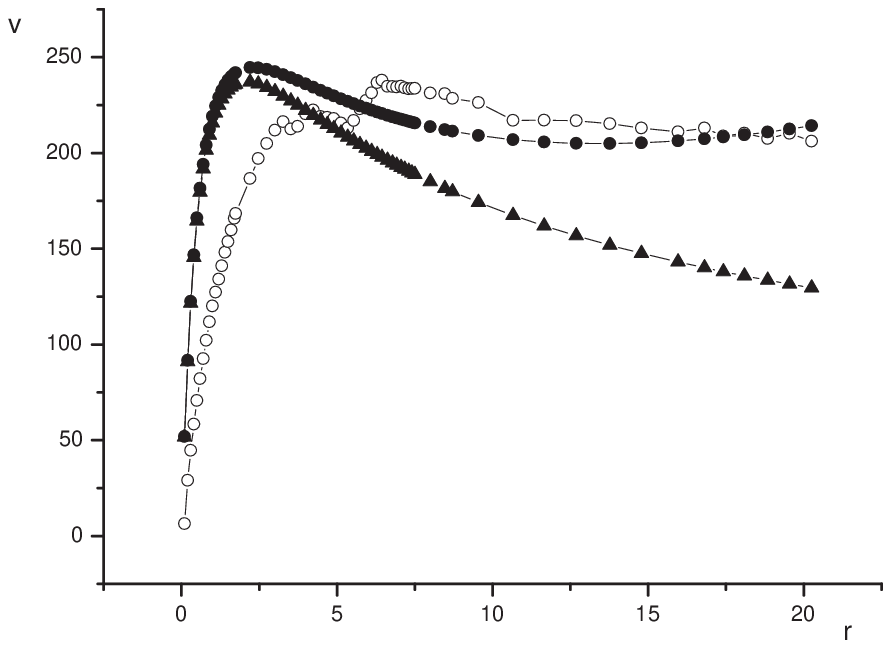}
%\caption{Ddo154; C=1.6} % Подпись рисунка
%\label{101} % Метка для ссылки на рисунок.
Fig.3. M 33; $\langle C\rangle=2.59*10^{-16}$.  \  \  \  \  \  \  \  \  \  \  \  \  \  \  \  \
Fig.4. Milky Way; $\langle C\rangle=1.89*10^{-16}$.
\end{figure}

\begin{figure}
 \hspace{-1.5cm}%   \centering
 \includegraphics[width=9cm,height=6.5cm]{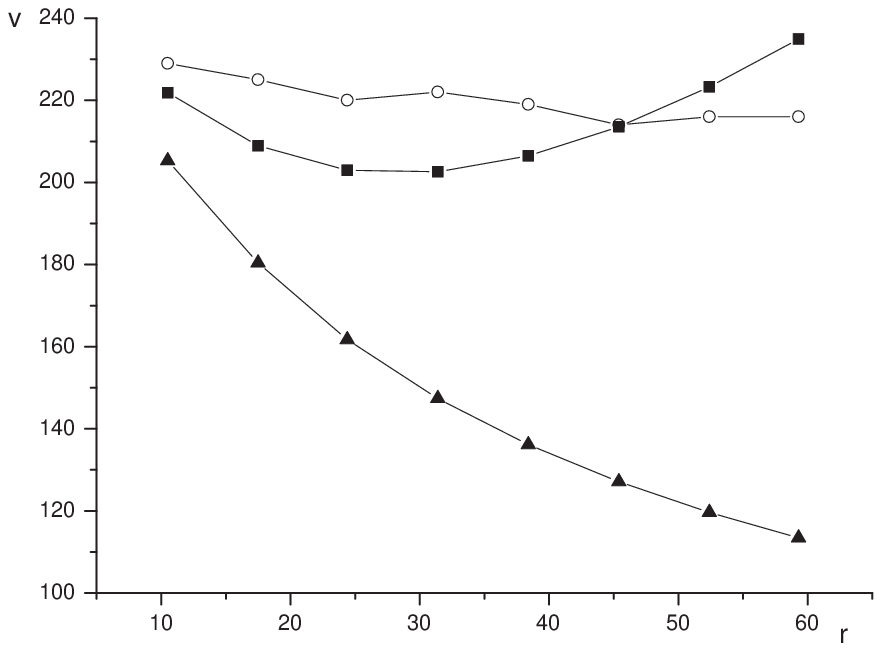}
\includegraphics[width=9cm,height=6.5cm]{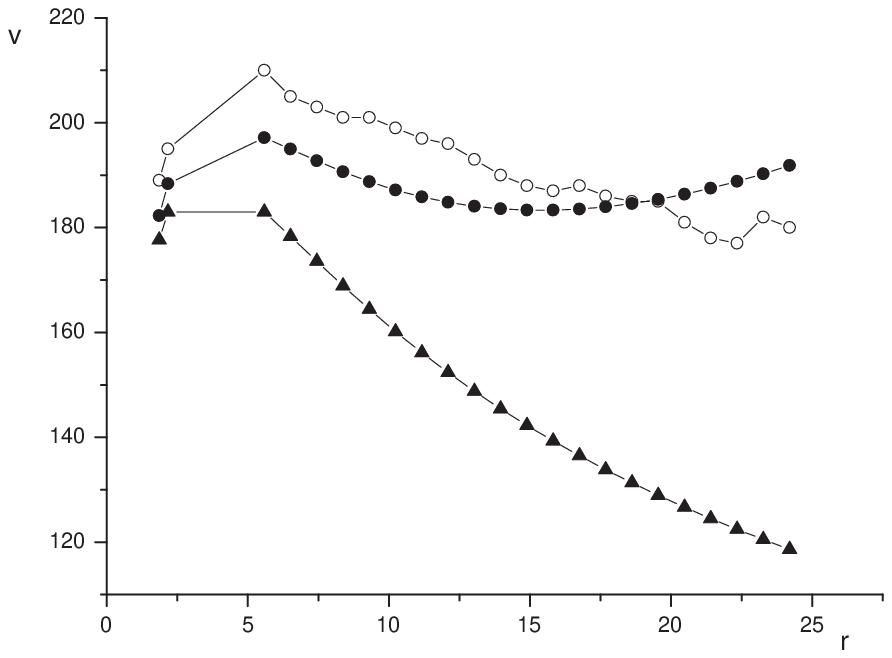}
Fig.5. NGC 801 ; $\langle C\rangle=9.865*10^{-17}$.  \  \  \  \  \  \  \  \  \  \  \
Fig.6. NGC 2903; $\langle C\rangle=1.59*10^{-16}$.
\end{figure}

\begin{figure}
 \hspace{-1.5cm}%   \centering
 \includegraphics[width=9cm,height=6.5cm]{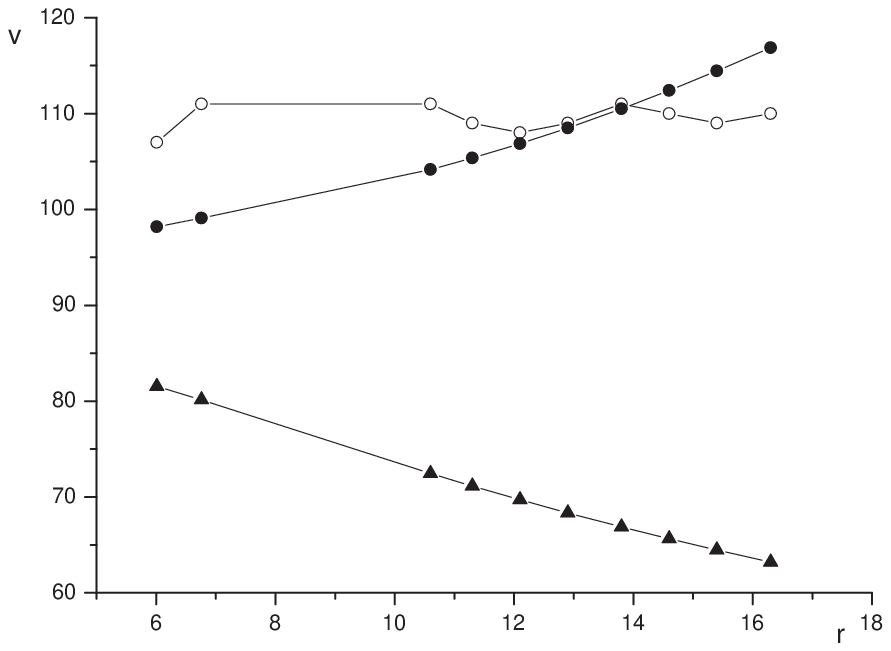}
\includegraphics[width=9cm,height=6.5cm]{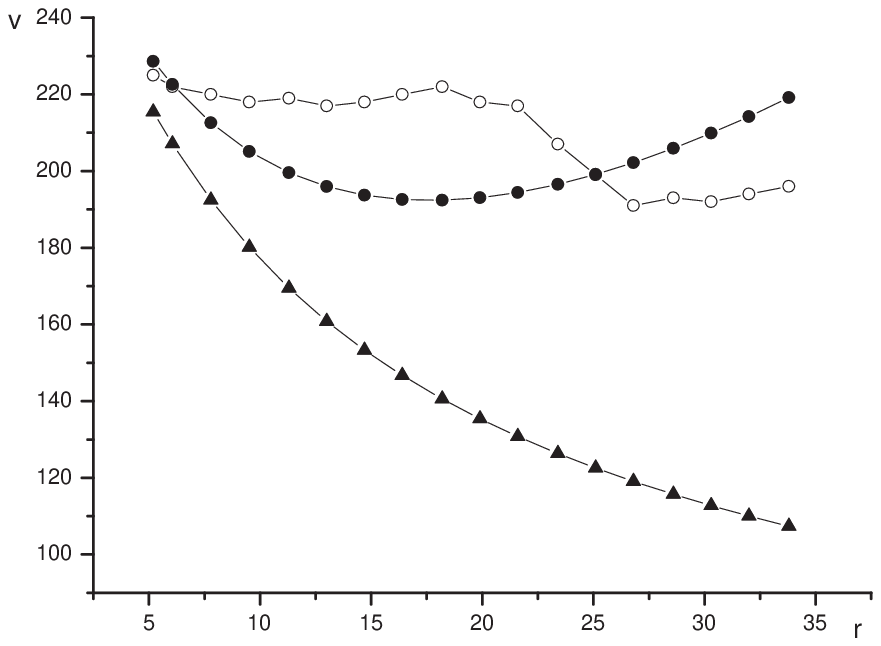}
Fig.7. NGC 4183; $\langle C\rangle=1.647*10^{-16}$.  \  \  \  \  \  \  \  \  \  \  \
Fig.8. NGC 5033; $\langle C\rangle=1.6*10^{-16}$.
\end{figure}

\begin{figure}
 \hspace{-1.5cm}%   \centering
 \includegraphics[width=9cm,height=6.5cm]{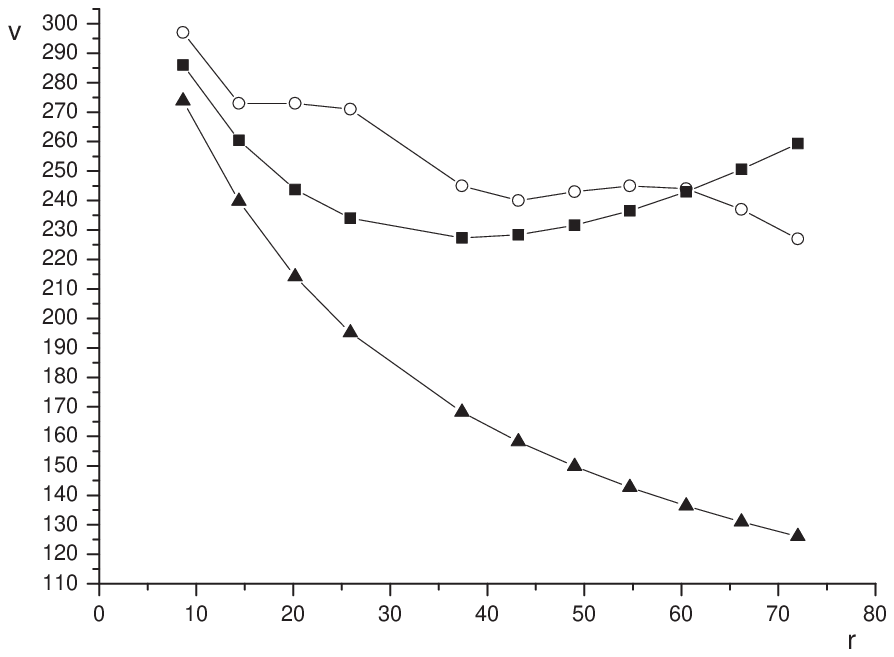}
\includegraphics[width=9cm,height=6.5cm]{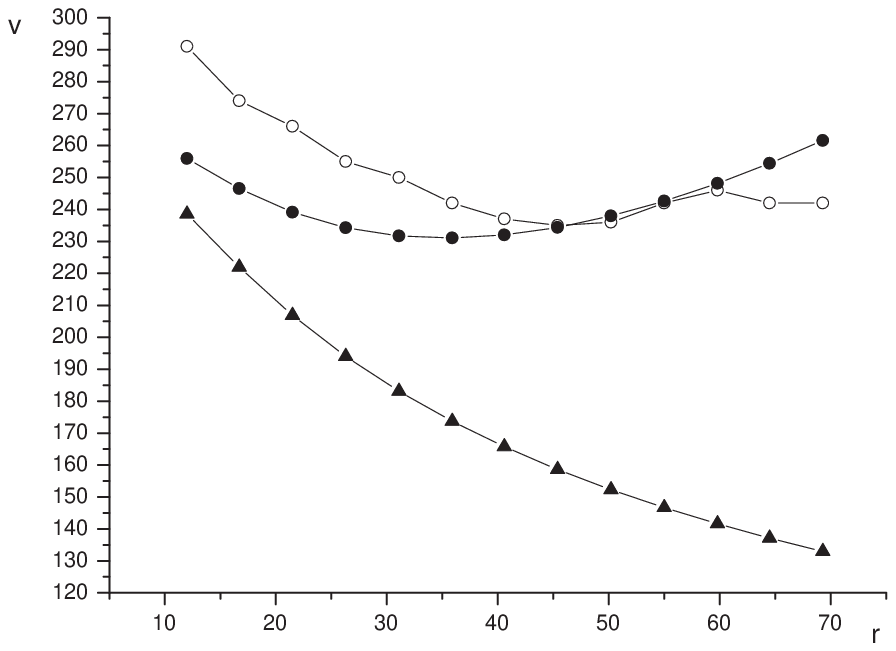}
Fig.9. NGC 5533; $\langle C\rangle=2.59*10^{-16}$.  \  \  \  \  \  \  \  \  \  \  \
Fig.10. NGC 6674; $\langle C\rangle=9.087*10^{-17}$.
\end{figure}

\begin{figure}
 \hspace{-1.5cm}%   \centering
 \includegraphics[width=9cm,height=6.5cm]{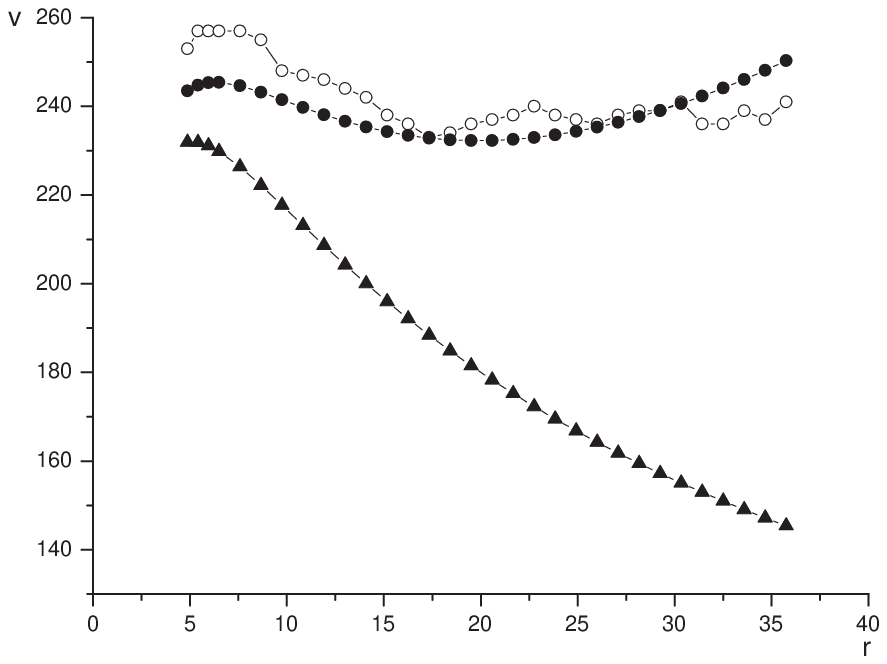}
\includegraphics[width=9cm,height=6.5cm]{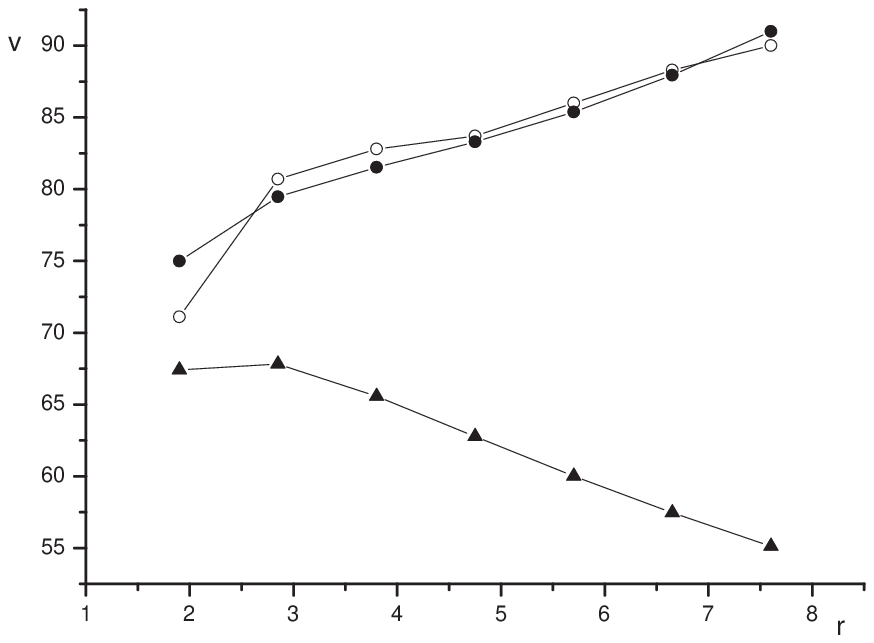}
Fig.11. NGC 7331; $\langle C\rangle=3.2*10^{-16}$.  \  \  \  \  \  \  \  \  \  \  \
Fig.12. UGC 2259; $\langle C\rangle=2.46*10^{-16}$.
\end{figure}

\begin{figure}
 \hspace{-1.5cm}%   \centering
 \includegraphics[width=9cm,height=6.5cm]{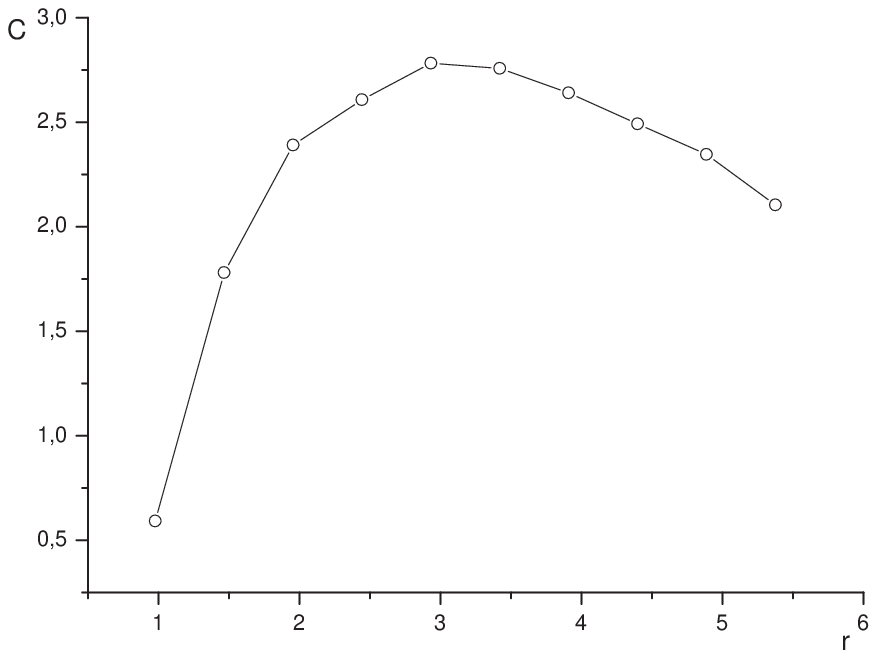}
\includegraphics[width=9cm,height=6.5cm]{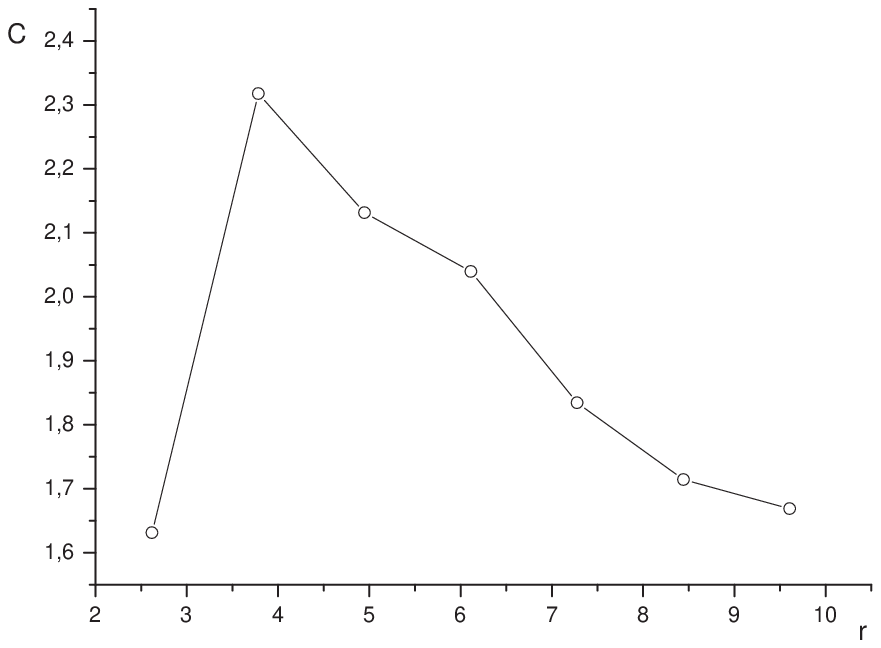}
Fig.13. DDO 154;   \  \  \  \  \  \  \  \  \  \  \  \  \  \  \  \  \  \  \  \  \  \  \  \  \  \  \  \  \  \  \  \  \  \  \  \  \  \
Fig.14. DDO 170;
\end{figure}

\begin{figure}
 \hspace{-1.5cm}%   \centering
 \includegraphics[width=9cm,height=6.5cm]{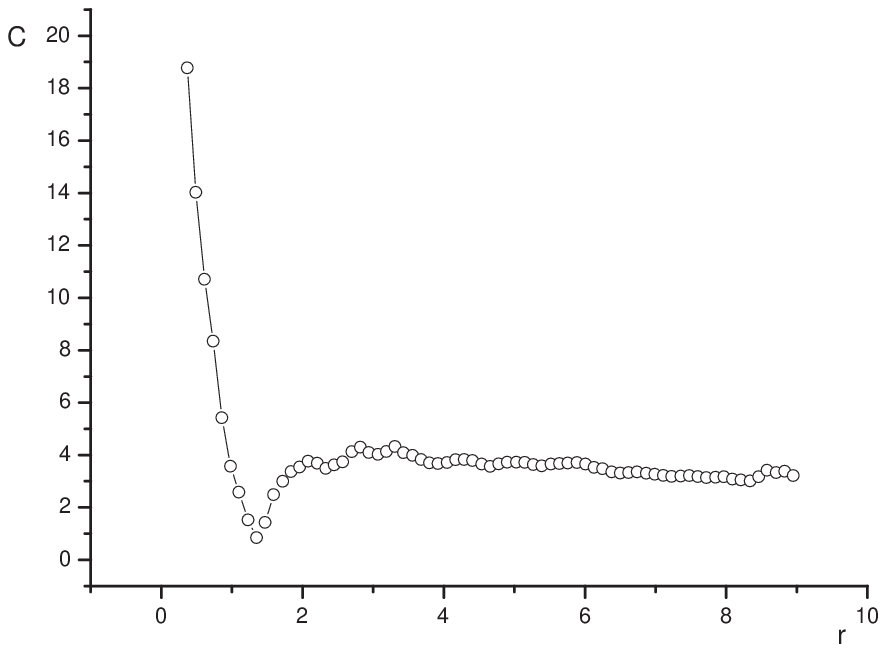}
\includegraphics[width=9cm,height=6.5cm]{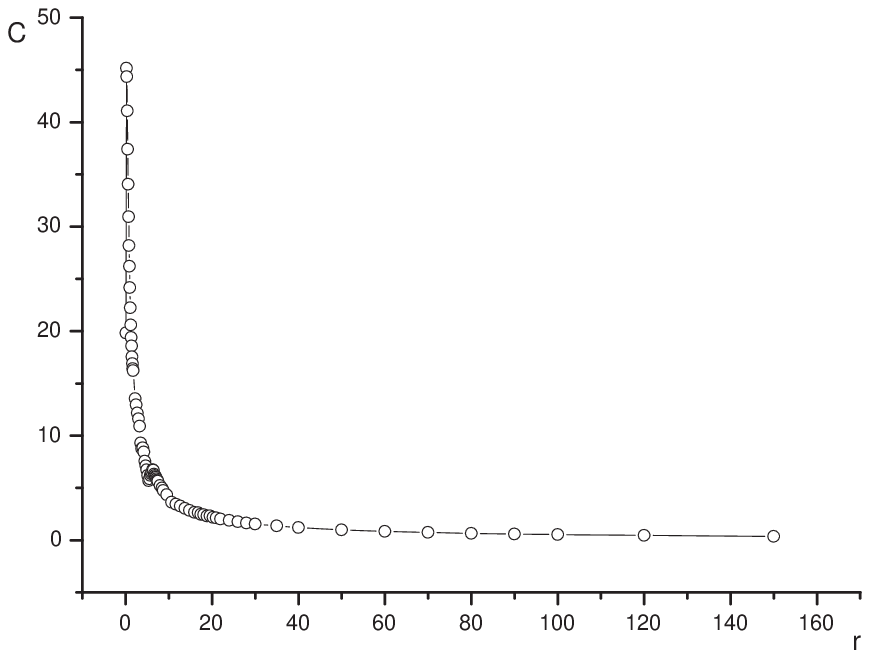}
Fig.15. M 33;   \  \  \  \  \  \  \  \  \  \  \  \  \  \  \  \  \  \  \  \  \  \  \  \  \  \  \  \  \  \  \  \  \  \  \  \  \  \  \  \  \  \  \  \
Fig.16. Milky Way;
\end{figure}

\begin{figure}
 \hspace{-1.5cm}%   \centering
 \includegraphics[width=9cm,height=6.5cm]{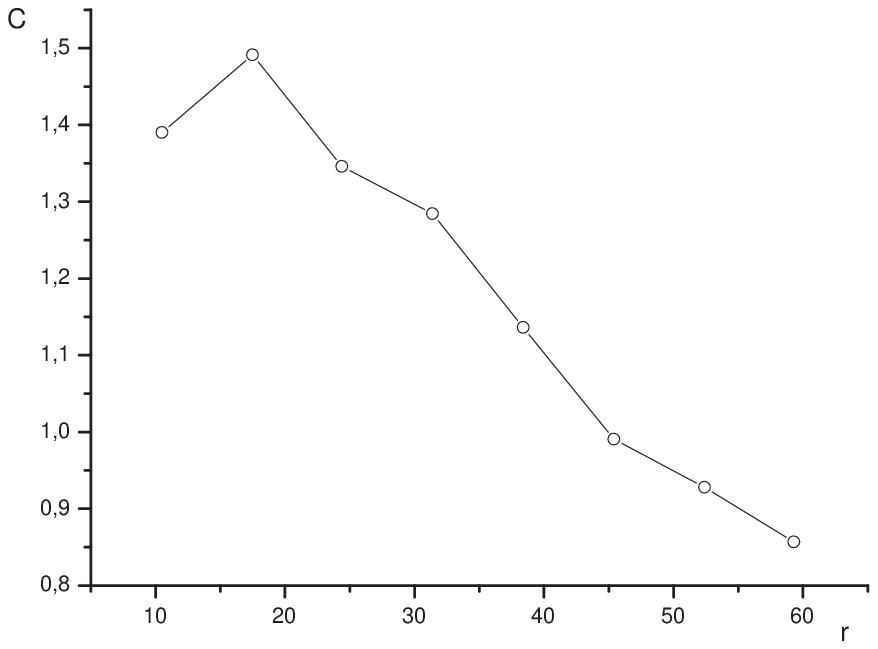}
\includegraphics[width=9cm,height=6.5cm]{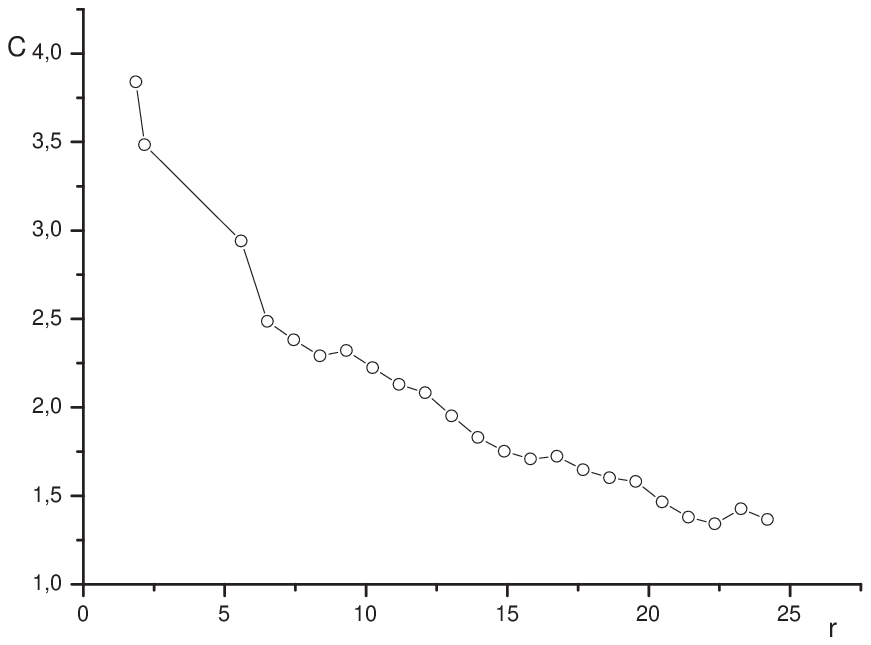}
Fig.17. NGC 801;   \  \  \  \  \  \  \  \  \  \  \  \  \  \  \  \  \  \  \  \  \  \  \  \  \  \  \  \  \  \  \  \  \  \  \  \  \  \  \
Fig.18. NGC 2903;
\end{figure}

\begin{figure}
 \hspace{-1.5cm}%   \centering
 \includegraphics[width=9cm,height=6.5cm]{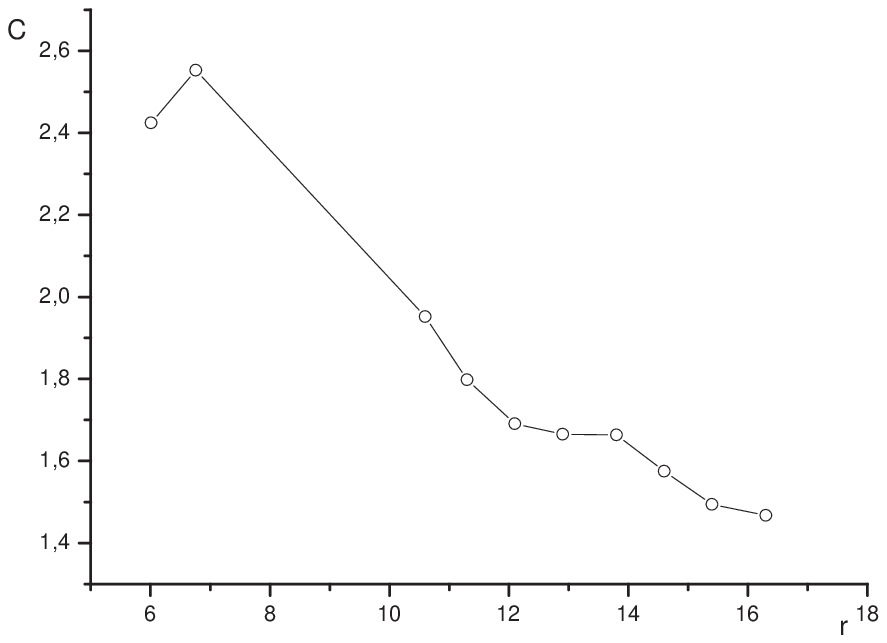}
\includegraphics[width=9cm,height=6.5cm]{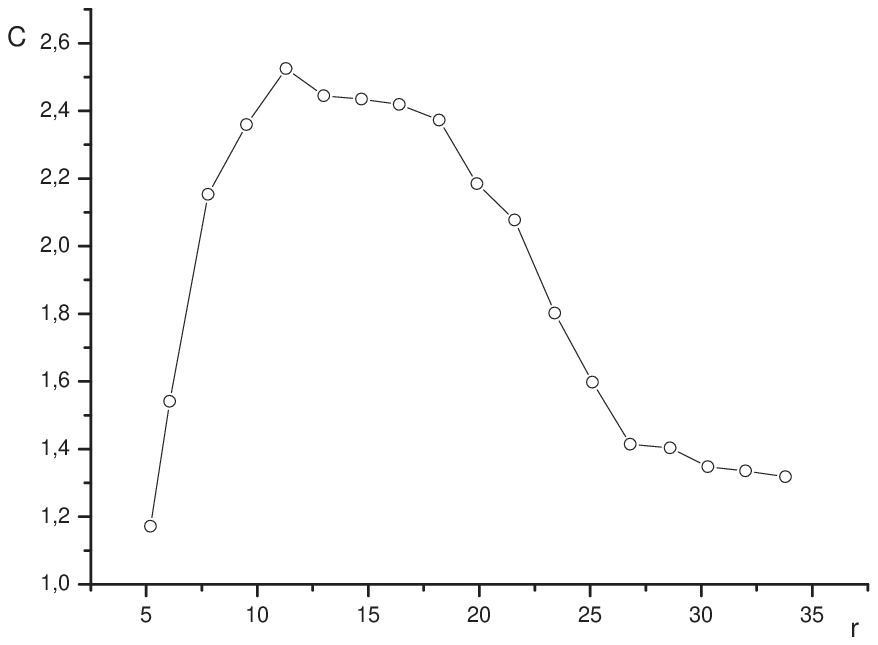}
Fig.19. NGC 4183;   \  \  \  \  \  \  \  \  \  \  \  \  \  \  \  \  \  \  \  \  \  \  \  \  \  \  \  \  \  \  \  \  \  \  \  \  \  \  \
Fig.20. NGC 5033;
\end{figure}

\begin{figure}
 \hspace{-1.5cm}%   \centering
 \includegraphics[width=9cm,height=6.5cm]{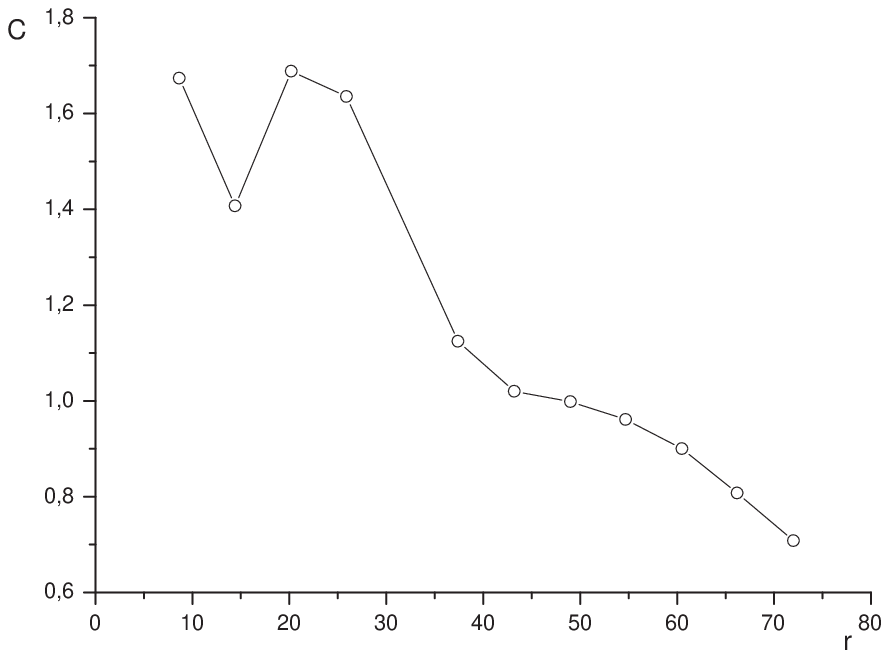}
\includegraphics[width=9cm,height=6.5cm]{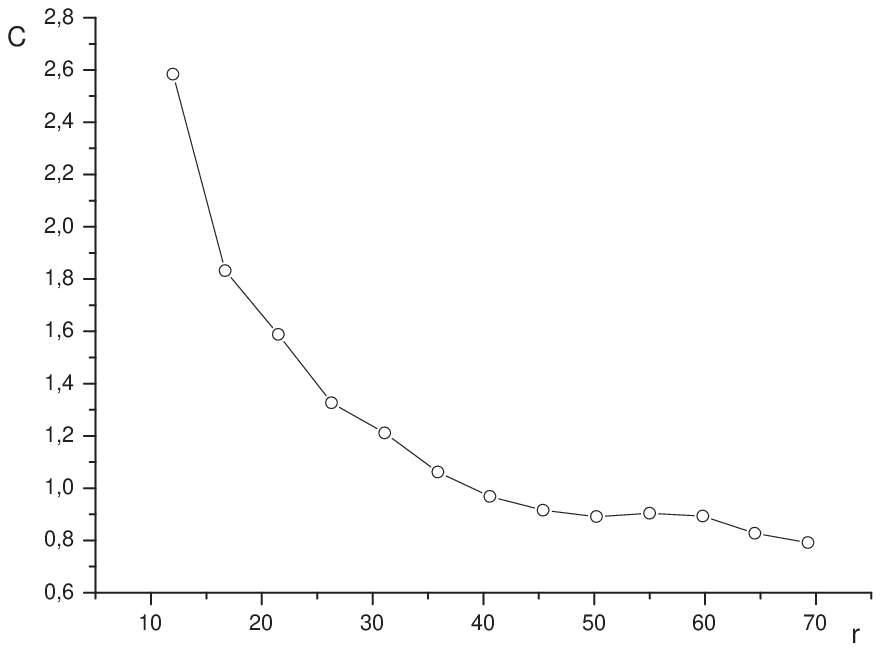}
Fig.21. NGC 5533;   \  \  \  \  \  \  \  \  \  \  \  \  \  \  \  \  \  \  \  \  \  \  \  \  \  \  \  \  \  \  \  \  \  \  \  \  \  \  \
Fig.22. NGC 6674;
\end{figure}

\begin{figure}
 \hspace{-1.5cm}%   \centering
 \includegraphics[width=9cm,height=6.5cm]{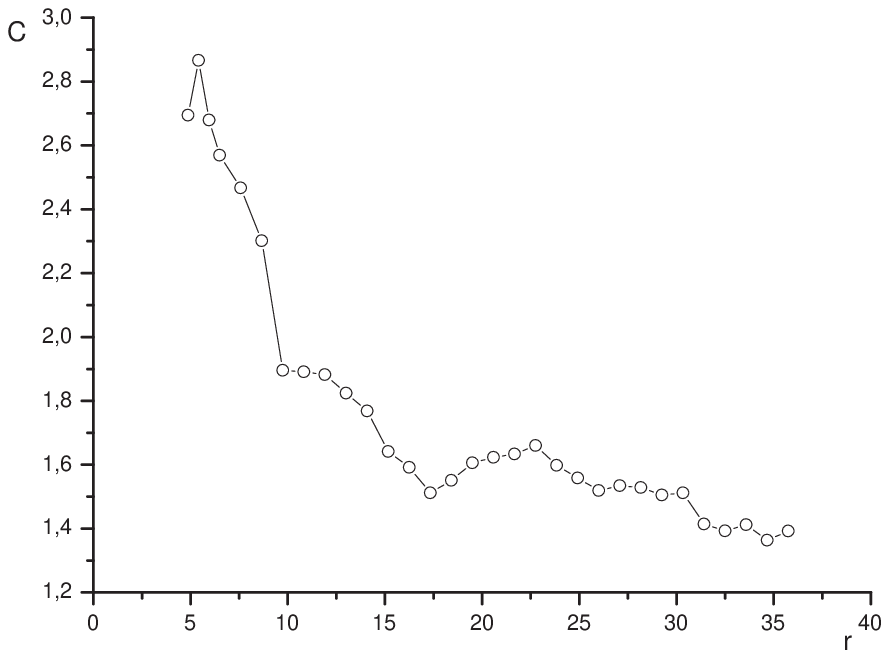}
\includegraphics[width=9cm,height=6.5cm]{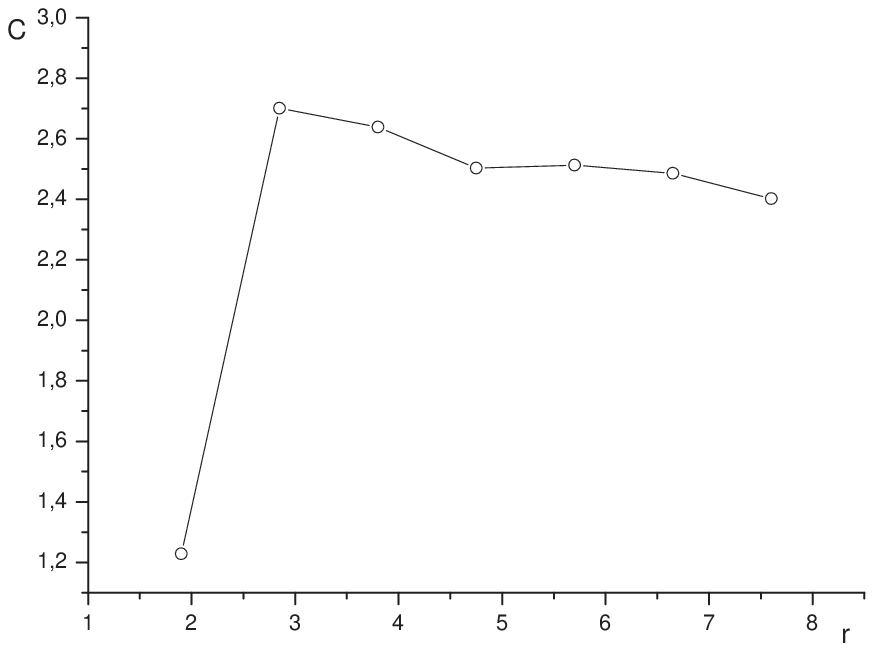}
Fig.23. NGC 7331;   \  \  \  \  \  \  \  \  \  \  \  \  \  \  \  \  \  \  \  \  \  \  \  \  \  \  \  \  \  \  \  \  \  \  \  \  \  \  \
Fig.24. UGC 2259;
\end{figure}

\begin{table}
\caption{Galaxy Properties}

\begin{center}
\begin{tabular}{|c|c|c|c|} \hline
 Galaxy  & Surface Brightness & Total mass M ($10^{10}M_{sun}$)\cite{5} & $r_c$ (kpc) \cite{5}  \\  \hline
 DDO 154 &       LSB          &   $0.13$ &   $0.53$            \\
 DDO 170 &       LSB          & $0.4$    &  $0.82$       \\
 M 33   &       LSB          & $0.93$    &  $0.6$       \\
 Milky Way &     HSB          & $9.12$    &  $1.04$       \\
 NGC 801 &     HSB          & $20.07$    &  $2.65$       \\
 NGC 2903 &     HSB          & $9.66$    &  $1.72$       \\
 NGC 4183 &     LSB          & $2.04$    &  $0.85$       \\
 NGC 5033 &     HSB          & $9.9$    &  $1.1$       \\
 NGC 5533 &     HSB          & $28.81$    &  $2.11$       \\
 NGC 6674 &     HSB          & $32.48$    &  $3.27$       \\
 NGC 7331 &     HSB          & $21.47$    &  $2.56$       \\
 UGC 2259 &     LSB          & $0.77$    &  $0.48$       \\  \hline
\end{tabular}
\end{center}
\end{table}

\section{Galaxy Rotation Curves. Exact formula.}

In the previous section, we have obtained the approximation of rotary curves of different
 galaxies on the assumption of existing of a medium constant cyclic field in the galaxies.
  It is seen that although approximation inclusive of a constant cyclic field considerably
   improves the Newtonian approximation, it basically has a qualitative character. The value
    of a medium cyclic field can be roughly connected with the average period of stars revolution $T$ in galaxies
$C\sim 2\pi/T$.

To get a precise coincidence of the curves, it is necessary to know the type of function
 of cyclic field distribution in each specific galaxy. The type of such a curve for different
  groups of similar galaxies can be someway selected, but using (\ref{30}) we can get
   the exact distribution of cyclic fields with the knowledge of rotation velocities by formula:

\begin{equation}\label{36}
C=\frac{v}{r}-\frac{\gamma m}{vr^2}
\end{equation}

In fig.13. - fig.24. there are graphs of distribution of the cyclic fields in the same
 galaxies for which the approximation with regard to the medium constant cyclic field
  was given earlier. Thus, using values of the cyclic fields (data should be multiplied by $10^{-16}$)
presented in the graphs we can get the perfect coincidence of theoretical and test values of
 the rotation curves. Therefore, in the framework of a vector theory of gravitation  \cite{6}
 the type of rotary curves of the galaxies can be explained directly without necessity
  of introduction of the additional hypothesis of unknown dark matter.

%%%%%%%%%%%%%%     Литература    %%%%%%%%%%%%%%%%%%%%%%%


\begin{thebibliography}{99}

\bibitem{1}
{\it J.P. Ostiker, P.J.E. Peebles, and A. Yahil.} ApJ, 193 (1974).
\bibitem{2}
{\it P.D. Sackett. et. al.} Nature. 370, 441, (1994).
\bibitem{3}
{\it E. Aprile. et. al.} arXiv:1104.2549 [astro-ph.CO].
\bibitem{4}
{\it M. Milgrom.} ApJ, 270, 365 (1983).
\bibitem{5}
{\it J. R. Brownstein, J. W. Moffat} ApJ, 636, 721 (2006); astro-ph/0506370.
\bibitem{6}
{\it V.N. Borodikhin.} Grav. Cosmol. 17, 161 (2011). arXiv:0802.2381 [gr-gq].
\bibitem{7}
{\it W.J.G. de Blok, S.S. McGaugh, and V.C. Rubin.}   AJ, 122, 2396  (2001).
\bibitem{8}
{\it S.S. McGaugh. et. al.} AJ, 659, 149 (2007).


\end{thebibliography}
\end{document}